\documentclass[aps,prb,showpacs,twocolumn,amsmath,amssymb,superscriptaddress,letterpaper]{revtex4}
\usepackage{times}
\usepackage{amsfonts}
\usepackage{mathrsfs}
\usepackage{graphicx}
\usepackage{dcolumn}
\usepackage{bm}
\usepackage{color}

\usepackage[colorlinks,bookmarks=false,citecolor=blue,linkcolor=red,urlcolor=blue]{hyperref}
\def\be{\begin{equation}}       \def\ee{\end{equation}}
\def\bea{\begin{eqnarray}}      \def\eea{\end{eqnarray}}

\begin{document}

\title{Quasi 1D topological nodal vortex line phase in doped superconducting 3D Dirac Semimetals}

\author{Shengshan Qin}
\affiliation{Kavli Institute of Theoretical Sciences, University of Chinese Academy of Sciences, Beijing, 100049, China}
\affiliation{Beijing National Research Center for Condensed Matter Physics, and Institute of Physics, Chinese Academy of Sciences, Beijing 100190, China}

\author{Lunhui Hu}
\affiliation{Department of Physics, University of California, San Diego, California 92093, USA}
\affiliation{Kavli Institute of Theoretical Sciences, University of Chinese Academy of Sciences, Beijing, 100049, China}

\author{Congcong Le}
\affiliation{Kavli Institute of Theoretical Sciences, University of Chinese Academy of Sciences, Beijing, 100049, China}
\affiliation{Beijing National Research Center for Condensed Matter Physics, and Institute of Physics, Chinese Academy of Sciences, Beijing 100190, China}

\author{Jinfeng Zeng}
\affiliation{Beijing National Research Center for Condensed Matter Physics, and Institute of Physics, Chinese Academy of Sciences, Beijing 100190, China}
\affiliation{University of Chinese Academy of Science, Beijing 100049, China}

\author{Fu-chun Zhang}
\affiliation{Kavli Institute of Theoretical Sciences, University of Chinese Academy of Sciences, Beijing, 100049, China}
\affiliation{CAS Center for Excellence in Topological Quantum Computation, University of Chinese Academy of Sciences, Beijing, 100049, China}

\author{Chen Fang}\email{cfang@iphy.ac.cn}
\affiliation{Beijing National Research Center for Condensed Matter Physics,
and Institute of Physics, Chinese Academy of Sciences, Beijing 100190, China}
\affiliation{Kavli Institute of Theoretical Sciences, University of Chinese Academy of Sciences, Beijing, 100049, China}
\affiliation{CAS Center for Excellence in Topological Quantum Computation, University of Chinese Academy of Sciences, Beijing, 100049, China}

\author{Jiangping Hu}\email{jphu@iphy.ac.cn}
\affiliation{Beijing National Research Center for Condensed Matter Physics,
and Institute of Physics, Chinese Academy of Sciences, Beijing 100190, China}
\affiliation{Kavli Institute of Theoretical Sciences, University of Chinese Academy of Sciences, Beijing, 100049, China}
\affiliation{CAS Center for Excellence in Topological Quantum Computation, University of Chinese Academy of Sciences, Beijing, 100049, China}
\affiliation{Collaborative Innovation Center of Quantum Matter, Beijing, China}

\date{\today}

\begin{abstract}
We study the vortex bound states in three dimensional (3D) superconducting Dirac semimetals with time reversal symmetry. Assuming two Dirac points on the $k_z$-axis and bulk $s$-wave superconductivity, with a quantum vortex line parallel to the $z$-direction, we find that the superconducting vortex line has a robust quasi-1D nodal phase. The nodal phase stems from the symmetry protected Dirac points in the normal state bands, and it can be characterized by a topological index $({\bf \nu}; {\bf n})$ at $k_z=0$ and $k_z=\pi$, where ${\bf \nu}$ is the $Z_2$ topological invariant for a 0D class-$D$ system and ${\bf n}$ is the $Z$ topological invariant for a 0D class-$A$ system according to the Altland-Zirnbauer classification. Based on the topological index, we find that vortex end Majorana zero mode can coexist with the quasi-1D nodal phase in certain kinds of Dirac semimetals. The influence of the symmetry breaking perturbations on the quasi-1D nodal phase is also analyzed. Finally, we discuss the possible material realization of such nodal vortex line state.
\end{abstract}



\maketitle

\emph{Introduction}.---Vortex bound states (VBSs) in superconducting topological materials have received great research interests in the past decade\cite{quantum_computation,TSC_Fu,TSC_edge2,TSC_edge3,Chiral_Zhang,Sau2010,Vishwanath_vortex,Huge_vortex1,Huge_vortex2,Dirac_equation1,Dirac_equation2,Fang_vortex,TSC_XuG,TSC_Liu1,TSC_Liu2}. Firstly proposed by Fu \emph{et al.}\cite{TSC_Fu}, the topologically protected surface Dirac cone of a 3D strong topological insulator (STI) proximity with a $s$-wave superconductor (SC), can result in Majorana zero mode (MZM) in the vortex core. This is very different from the condition of a conventional SC, where the VBSs are the Caroli-de Gennes-Matricon excitations with a tiny gap proportional to $\Delta^2/E_f$\cite{Caroli}, where $\Delta$ is the superconducting order parameter and $E_f$ is the Fermi energy. Later on, P. Hosur \emph{et al.}\cite{Vishwanath_vortex} point out that the above vortex bound MZMs can be generally understood in the viewpoint of the topological phase transition of the VBSs, the so-called vortex phase transition (VPT), if the 3DSTI can be doped to be superconducting. In recent years, more and more signatures for the existence of vortex bound MZMs have been observed in experiments\cite{TSC_Jia1,TSC_Jia,TSC_iron1,TSC_iron2,FeSe_feng,TSC_iron3,PdBi2_xue}.

Besides the TIs\cite{TSC_edge4,2DTI_Kane1,2DTI_Kane2,2DTI_Kane3,2DTI_Zhang,3DTI}, there is also a class of topological metallic systems, the topological semimetals (TSMs)\cite{TSC_edge1,TSM_rev,WS_review} including Dirac semimetals\cite{Dirac Kane,Na3Bi,Cr3As2} (DSs) and Weyl semimetals\cite{HgCrSe,XG Wang,multilayer weyl,nodal Balents,TRS_WS Balents,TaAs thory1,TaAs thory2}, which usually emerge as intermediate states between the TIs and the normal insulators (NIs)\cite{intermediate1,intermediate2}. In recent years, many TSMs have been found to have bulk superconductivity\cite{SC_SM1,SC_SM2,SC_SM3,IrTe2_SC1,IrTe2_SC2}. Therefore, it is natural to ask what the condition is for the VBSs in a superconducting TSM. The question is partly answered for the TWSs without time reversal symmetry (TRS), which turns out to have flat bands in the superconducting vortex line\cite{vortex_flat}. However, a complete comprehension still lacks.

In this letter, we study the VBSs in superconducting DSs with TRS, and develop a general theory to describe the topological VPTs. Generally, 3D DSs can be classified into topologically trivial and nontrivial ones according to the $Z_2^{TI}$ topological index in the time reversal invariant (TRI) planes in the 3D Brillouin zone\cite{DS_Nagosa}, which will be referred as trivial and nontrivial DSs in the following. We take the DSs, both the trivial and nontrivial cases, protected by the $C_{4z}$-rotational symmetry in Ref.\cite{DS_Nagosa} for instance, and carry out numerical and analytical analysis on the topological VPTs in the weak pairing limit by assuming bulk $s$-wave superconductivity and a single straight quantum vortex line along the $k_z$-axis. Our main results are summarized as follows. (i) For both kinds of DSs, the energy spectrum of the superconducting vortex line has pairs of nodal points on the $k_z$-axis for certain range of doping level, namely there is a robust gapless phase. The nodal phase is closely related to the symmetry protected Dirac points in the normal state band structures. (ii) However, the VPTs are very different for the two kinds of DSs at the TRI momenta: there are two (one) VPTs at $k_z=0$ for the trivial (nontrivial) DSs by tuning the chemical potential from the Dirac points, while there can never be such process at $k_z=\pi$ for both kinds of DSs. We prove that the number of the VPTs at the TRI momenta is determined by the $Z_2^{TI}$ topological index of the normal state band structures in the corresponding TRI planes. (iii) The topological VPTs in both kinds of DSs can be well depicted by a topological index $({\bf \nu}; {\bf n})$, where ${\bf \nu}$ is the $Z_2^{0D}$ topological invariant for a 0D class-$D$ system and ${\bf n}$ is the $Z^{0D}$ topological invariant for a 0D class-$A$ system according to the Altland-Zirnbauer (AZ) classification\cite{TSC_classification1,TSC_classification2}. Interestingly, based on the index it can be inferred that the vortex end MZM can exist even in a gapless system, and such a state can be realized in the trivial DSs. Finally, we analyze the influence of the symmetry breaking perturbations on the topological VPTs, and discuss the material realization of the nodal superconducting vortex line state.

We give a general description for the problem at first. For a 3D SC which has $s$-wave spin-singlet superconductivity, its Hamiltonian in the basis $\Psi^\dag({\bf k})=(c_{\uparrow}^\dag({\bf k}),c_{\downarrow}^\dag({\bf k}),c_{\uparrow}(-{\bf k}),c_{\downarrow}(-{\bf k}))$ has the following form
\begin{eqnarray}\label{Hsc}
H_{sc}&=&\left(\begin{array}{cc}
          H_0(\mathbf{k})-\mu     &     \Delta                        \\
          \Delta^\dagger          &     \mu-H_0^\ast(-\mathbf{k})     \\
          \end{array}
          \right),
\end{eqnarray}
where only the indices for the spin and Nambu spaces have been preserved. In the equation, $\Delta$ is the superconducting order parameter which takes the form $\Delta_0i\sigma_2$ in the spin space, $\mu$  is the chemical potential, and $H_0(\mathbf{k})$  is the normal state Hamiltonian. If a single quantum vortex goes through the SC along the $z$-direction, the superconducting order parameter $\Delta(\mathbf{r})$ in the real space is  transformed into $\Delta(r)e^{i\theta}$, where $r=\sqrt{x^2+y^2}$ and $\theta$ is the polar angle\cite{Vishwanath_vortex,Fang_vortex}. Obviously, the vortex line destroys the translational symmetry in the $xy$-plane. The TRS is broken while the particle-hole symmetry (PHS) is preserved. Therefore, the 3D SC with a vortex line can be viewed as a quasi-1D system belonging to the class-$D$ of the AZ classification\cite{TSC_classification1,TSC_classification2}, whose topological property is characterized by its fermion parity for a full-gap system.

%
%


\begin{figure*}
\centerline{\includegraphics[width=0.95\textwidth]{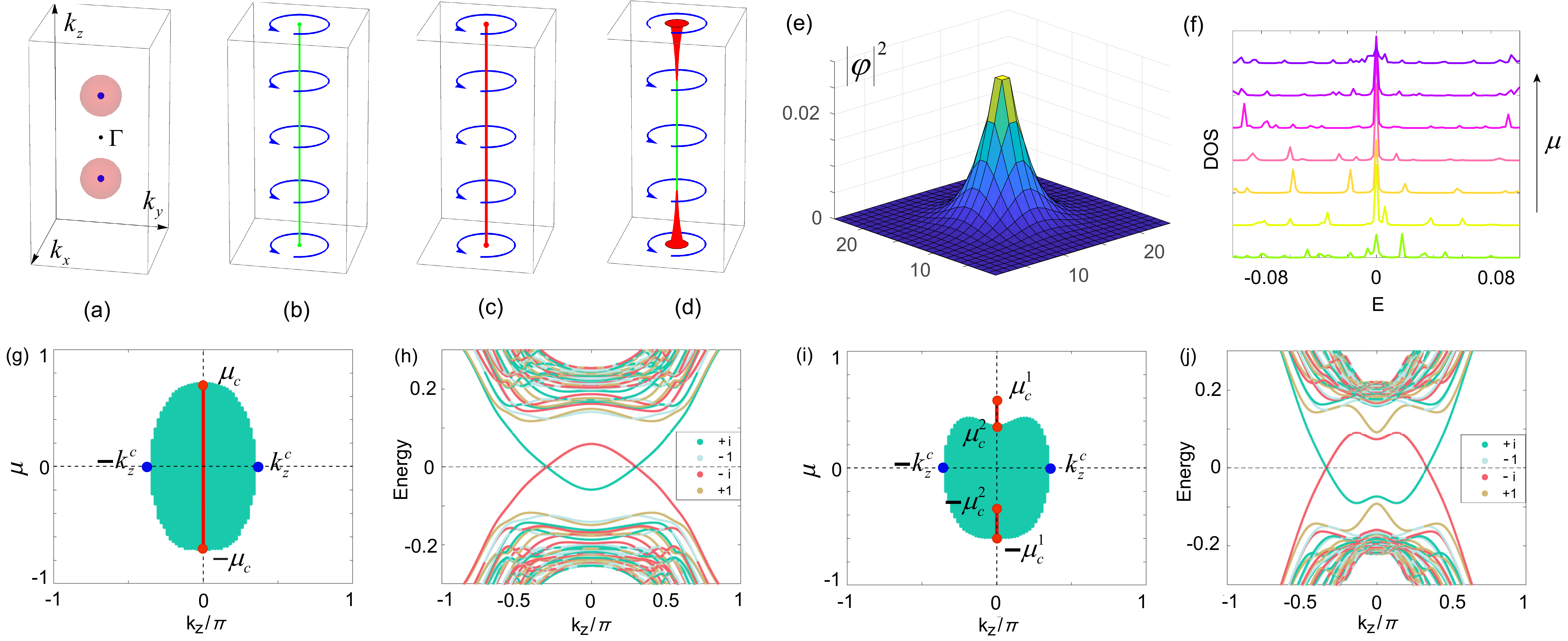}}
\caption{(color online) (a) A $A_{1g}$ superconducting order $\Delta_0i\Sigma_{02}$ opens up full gap on the FSs in doped DSs. (b)$\sim$(d) show sketch of different topological phases of the VBSs: (b) shows a full-gap topologically trivial state, (c) shows a topological nodal state and (d) shows a full-gap topologically nontrivial state with vortex end MZMs. (g) and (i) show the topological phase diagrams of the VBSs for the DSs in $H_0^{(1)}$ and $H_0^{(2)}$ respectively. The blue points here indicate the location of the Dirac points on the $k_z$-axis, and the edge of the green region represents the trajectory of the nodal points of the vortex line in the $(\mu, k_z)$ space. In the green region the $Z^{0D}$ topological index satisfies $Q=1$, and the $Z_2^{0D}$ topological invariant satisfies $\nu=-1$ on the red line at $k_z=0$. Specifically, the $Z_2^{0D}$ index in different rotational subspaces satisfies $\nu_{+1}=-1$ for $-\mu_c^2<\mu<\mu_c^1$ and $\nu_{+i\oplus -i}=-1$ for $-\mu_c^1<\mu<\mu_c^2$ in (i), while it only has nontrivial $\nu_{+i\oplus -i}$ for $-\mu_c<\mu<\mu_c$ in (g). The topologically nontrivial $\nu_{+1}$ in (i) can lead to vortex end MZM even though the whole vortex line is nodal, and (f) shows the density of states (DOS) at the end of the vortex line when $\mu=\{-0.2, -0.1, 0.0, 0.1, 0.2, 0.3, 0.35\}$. (h) and (j) show the low-energy spectrum of the vortex line in different rotational subspaces (represented by different colors) as a function $k_z$ at $\mu=0.44$ for the nontrivial DS and $\mu=0.20$ for the trivial DS respectively. Obviously, the nodal points on the $k_z$-axis is attributed to states form the $H_{+i}$ and $H_{-i}$ subspaces. (e) shows the distribution of the eigenfuction in the $xy$-plane corresponding to the zero mode on the $k_z$-axis in (h) and (j). In the calculation of (f) the lattice size is taken as $14\times14\times26$, while for other figures the lattice size in the $xy$-plane is $24\times 24$. The SC order parameter is set to be $\Delta_0=0.2$ and the vortex size is supposed to be infinite small. Without special instructions, the lattice and SC parameters in the following calculations are all the same with that in Fig.\ref{vortex_DS}.
\label{vortex_DS}}
\end{figure*}

\emph{numerical results}.---We study the VBSs in doped $s$-wave superconducting DSs numerically, including both the trivial and nontrivial ones, based on the models with $C_{4z}$-rotational symmetry in Ref.\cite{DS_Nagosa}. In the calculations, we take a simple form of $\Delta(r)$: $\Delta(r)=\Delta_0\Theta(r-R)$, where $\Theta(r)$ is the step function and $R$ is the vortex size.

We start with the nontrivial DS, whose normal state Hamiltonian reads as
\begin{eqnarray}\label{H_TDS}
H_0^{(1)}&=&(m-t\cos k_x-t\cos k_y-t_3\cos k_z)\Sigma_{30}  \nonumber\\
&+&t^\prime \sin k_x\Sigma_{13}+t_3^\prime \sin k_z(\cos k_x-\cos k_y)\Sigma_{11}  \nonumber\\
&-&t^\prime \sin k_y\Sigma_{20}+2t_3^\prime \sin k_z\sin k_x\sin k_y\Sigma_{12},
\end{eqnarray}
where $\Sigma_{ij}=\tau_i\sigma_j$, and $\tau_i$ and $\sigma_i$ ($i=0, 1, 2, 3$) are the Pauli matrices representing the orbital and spin degrees of freedom. Remarkably, this Hamiltonian is equivalent to the $k\cdot p$ model for Cd$_3$As$_2$\cite{Cr3As2}, which respects the TRS and the full symmetry of the $D_{4h}$ point group. The basis of the model contains two Kramers' doublets with opposite parity, whose angular momentum $J_z$ defined according to the $C_{4z}$-rotational axis are $\pm\frac{1}{2}$ and $\pm\frac{3}{2}$ respectively. The related symmetry group generators are given by the following matrices: $T=i\Sigma_{02}K$, $I=\Sigma_{30}$, $C_{4z}=(\Sigma_{30}+i\Sigma_{03})/\sqrt{2}$ and $C_{2x}=\Sigma_{31}$. We set the parameters as $\{t, t_3, t^\prime, t_3^\prime, m\}=\{1.0, 0.5, 1.0, 1.0, 2.2\}$ to derive a TDS state, where $H_0^{(1)}$ has a band inversion at the $\Gamma$ point and two Dirac points at $\pm Q=(0,0,\pm k_z^c)$, where $k_z^c=\arccos0.4$. We obtain the VBSs by diagonalizing the Hamiltonian directly. We first focus on the gap-close-reopen process at the TRI momenta, and find that such a process occurs at $k_z=0$ at $\mu_c=0.71$, while it never happens at $k_z=\pi$, as shown in Fig.\ref{vortex_DS}(g). This is similar to the 3D STI case\cite{Vishwanath_vortex}. For the STIs, the superconducting vortex line has a topologically nontrivial full-gap phase\cite{Vishwanath_vortex} when $|\mu|<\mu_c$. However, there is no such phase here, instead it has a robust 1D nodal phase. Fig.\ref{vortex_DS}(h) shows that the vortex line has two nodal points at $k_z=\pm 0.3\pi$ when $\mu=0.44$. When the chemical potential is changed, the nodal points will move along the $k_z$-axis, and they can never gap out until the two nodal points meet with each other at $k_z=0$, as presented in Fig.\ref{vortex_DS}(g).


Then, we turn to the trivial DS whose normal state Hamiltonian reads as
\begin{eqnarray}\label{H_NDS}
H_0^{(2)}&=&(m-t\cos k_x-t\cos k_y-t_3\cos k_z)\Sigma_{30}  \nonumber\\
&+&t^\prime\sin k_z \sin k_x\Sigma_{10}+t_3^\prime (\cos k_x-\cos k_y)\Sigma_{22}  \nonumber\\
&-&t^\prime\sin k_z \sin k_y\Sigma_{23}+2t_3^\prime \sin k_x\sin k_y\Sigma_{21}.
\end{eqnarray}
Compared to $H_0^{(1)}$, the above Hamiltonian have the same symmetry group and its basis also consists of two Kramers' doublets with $J_z=\pm\frac{1}{2}$ and $J_z=\pm\frac{3}{2}$, but the two Kramers' doublets have the same parity here. Therefore, the IS operator has the form $I=\Sigma_{00}$ and the matrix form of the other symmetry operators maintain the same in this case. We set all the parameters the same as that in $H_0^{(1)}$, leading to that $H_0^{(2)}$ has the same band inversion and Dirac points with $H_0^{(1)}$. However, since the band inversion occurs between two Kramers' doublets with the same parity, $H_0^{(2)}$ describes a trivial DS\cite{Z2 inversion}. The topological property of the VBSs for such a trivial DS is even more exotic. It turns out that, the vortex line has a robust 1D nodal phase, which is similar to the nontrivial DS case. Fig.\ref{vortex_DS}(j) shows that, the vortex line has two nodal points at $k_z=\pm 0.34\pi$ when the chemical potential is $0.20$. However, the condition is very different at the TRI momenta. There are two gap-close-reopen processes at $\mu_c^1=0.594$ and $\mu_c^2=0.357$ at $k_z=0$, while it has no such process at $k_z=\pi$, as shown in Fig.\ref{vortex_DS}(i). We have also checked the evolution of the nodal points carefully, and find that: the nodal point which occurs at $\mu_c^1$ is directly gapped out, while the one at $\mu_c^2$ moves along the $k_z$-axis, if the chemical potential $\mu$ is changed.

\emph{General theory}.---We develop a general theory to account for the different vortex line phases in the two kinds of DSs with intrinsic $s$-wave superconductivity. We begin with the vortex line at $k_z=0/\pi$. Since the PHS preserves at the TRI momenta, the vortex line can be viewed as a quasi-0D system belonging to the class-$D$ of the AZ classification at $k_z=0/\pi$. The topological index for such a 0D system is $Z_2^{0D}$, which characterizes the fermion parity of the system. A gap-close-reopen process in such a 0D system means a topological phase transition, namely a topological VPT. In the normal state, $H_0^{(1)}$ describes a 2D TI (NI) in the $k_z=0$ ($k_z=\pi$) plane, while $H_0^{(2)}$ describes a 2D NI (NI) in the $k_z=0$ ($k_z=\pi$) plane. As we shall show, there is a correspondence between $Z_2^{0D}$ and $Z_2^{TI}$ in the weak pairing limit. We begin with a 2D TI without superconductivity and set the chemical potential in the insulating gap. If there is a $\pi$-flux, the ground state of such a finite-size 2D TI is doubly degenerate\cite{Z2 pump}. The $\pi$-flux for electrons means a quantum vortex in the superconducting state. As superconductivity is turned on, the superconducting order parameter will destroy the ground state and make one of the degenerate ground state occupied while the other one unoccupied. On the other hand, though the superconductivity breaks the $U(1)$ charge conservation symmetry, the charge (fermion) parity is still well-define. Immediately, we can concluded that the fermion parity of the quasi-0D vortex system must be odd, namely the quasi-0D system is $Z_2^{0D}$ nontrivial. Therefore, a correspondence between $Z_2^{0D}$ and $Z_2^{TI}$ is established, when the chemical potential is in the gap of the TI. Based on the above argument, it can be straightforwardly concluded: the vortex line corresponding to the DS in $H_0^{(1)}$ must go through odd (even) times of VPTs at $k_z=0$ ($k_z=\pi$), while it must go through even (even) times of VPTs at $k_z=0$ ($k_z=\pi$) for the DS in $H_0^{(2)}$, if the chemical potential is tuned from in the insulation gap to infinite large, since the system is always topologically trivial if the chemical potential is infinite large. This is consistent with the numerical results. However, we can not tell whether the vortex line is full-gap or not from the $Z_2^{0D}$ topological invariant.

%

Recalling that DSs are always protected by some rotational symmetry and the vortex line along the rotational axis does not break the rotational symmetry, we can also classify the superconducting vortex line according to the rotational symmetry\cite{rotation1,rotation2,rotation3}. For the above superconducting DSs with a vortex line, the $C_{4z}$-rotational symmetry has the following matrix form\cite{supplementary} in the basis in Eq.\ref{Hsc}
\begin{eqnarray}\label{C4_BdG_vortex}
C_{4z}^{vortex}=\delta_{ij}\left(
            \begin{array}{cc}
            C_{4z}e^{i\pi/4}             &                      0                      \\
            0                            &            C_{4z}^\ast e^{-i\pi/4}          \\
            \end{array}
          \right),
\end{eqnarray}
where $C_{4z}$ is the matrix form of the $C_{4z}$-rotational operator in the normal state, $i$ and $j$ are the site index in the $xy$-plane which are related with each other by the $C_{4z}$-rotational symmetry, and the phase factor $e^{\pm i\pi/4}$ stems from the $U(1)$ gauge field of the quantum vortex line. Obviously, the magnetic flux transforms the eigenvalues of the $C_{4z}$-rotational symmetry from $e^{i(2m-1)\pi/4}$ to $e^{im\pi/2}$, where $m$ is an integer. Straightforwardly, the superconducting vortex line system can be decoupled into four subspaces according to $C_{4z}^{vortex}$
\begin{eqnarray}\label{decouple}
H_{sc}=H_{+1}\oplus H_{+i}\oplus H_{-1}\oplus H_{-i},
\end{eqnarray}
where the subsripts are the four eigenvalues of the $C_{4z}$-rotational symmetry.

We can analyze the topological property in each subspace at $k_z=0/\pi$. Considering that the PHS is an anti-unitary symmetry, the four subspaces can be classified into two categories. For the two subspaces with real (imaginary) eigenvalues, the PHS preserves (breaks) within each subspace. Therefore, $H_{+1}$ and $H_{-1}$ ($H_{+i}$ and $H_{-i}$) belong to the 0D class-$D$ (class-$A$) of the AZ classification at the time reversal invariant momenta\cite{TSC_classification1,TSC_classification2}, which is characterized by the $Z_2^{0D}$ ($Z^{0D}$) topological invariant. However, since $H_{+i}$ is inverted into $H_{-i}$ under the PHS, the subspace $H_{+i}\oplus H_{-i}$ regains the PHS and belongs to the 0D class-$D$ of the AZ classification at the TRI momenta. As mentioned above, $Z_2^{0D}$ characterizes the fermion parity of each of the three PHS symmetric subspaces, while $Z^{0D}$ is defined as the number of the eigenstates with negative energy in each of the four rotational subspaces. Immediately, we can come to the following conclusions. (i) At the TRI momenta, the $Z_2^{0D}$ topological invariants $\nu$ for the total system and the three PHS symmetric subspaces have the relationship
\begin{eqnarray}\label{Z2}
\nu=\nu_{+1}\cdot\nu_{-1}\cdot\nu_{+i\oplus -i},
\end{eqnarray}
where $\nu=1$ ($\nu=-1$) corresponds to the topological trivial (nontrivial) state. (ii) If the $Z^{0D}$ topological invariant $N$ is different at $k_z=0$ and $k_z=\pi$, namely $N_\alpha(0)\neq N_\alpha(\pi)$, where $\alpha$ stands for the four eigenvalues of the $C_{4z}$-rotational symmetry, the vortex line must have a nodal phase. (iii) The change of $\nu_{+1}$ ($\nu_{-1}$) has no influence on the $N_{+1}$ ($N_{-1}$), while in the $H_{+i}\oplus H_{-i}$ subspace the $Z_2^{0D}$ and $Z^{0D}$ topological invariant has the relationship
\begin{eqnarray}\label{Z2&Z}
\nu_{+i\oplus -i}=e^{i\pi(N_{+i}-N_{-i})/2}.
\end{eqnarray}
(iv) At $k_z\neq 0/\pi$, the $Z^{0D}$ topological invariant is still well-defined while the $Z_2^{0D}$ topological index is no longer well-defined. Therefore, the VPTs in the $H_{+i}\oplus H_{-i}$ subspace at the time reversal invariant momenta can evolve into a 1D nodal phase, while the VPTs in the the $H_{+1}$ and $H_{-1}$ subspaces merely emerge as an accidental nodal point in the superconducting vortex line. Straightforwardly, it can be inferred that if both $\nu_{+1}$ ($\nu_{-1}$) and $\nu_{+i\oplus -i}$ are topologically nontrivial, there can exist rotational symmetry protected vortex end MZM even though the whole system is gapless.

The analysis can be directly applied to the above trivial and nontrivial DSs. Since there is no band inversion at $k_z=\pi$ for both of the DSs, the $Z^{0D}$ topological invariant are all the same in the four subspaces. Therefore, we can use the following topological number at $k_z=0$ to characterize the nodal vortex line phase
\begin{eqnarray}\label{Z phase}
Q=\frac{N_{+i}(0)-N_{-i}(0)}{2}.
\end{eqnarray}
For any $Q\neq 0$, there will be $Q$ pairs of nodal points on the $k_z$-axis. In our calculations, we find that the VPT at $\mu_c$ ($\mu_c^2$) at $k_z=0$ for the nontrivial (trivial) DS changes $Q$ from 0 to 1, leading to that there must be a nodal vortex line phase with one pair of nodal points on the $k_z$-axis\cite{supplementary}. The VPT at $\mu_c^1$ at $k_z=0$ for the trivial DS occurs in the $H_{+1}$ subspace, which leads to that there is vortex end MZM even though the whole vortex line is gapless, as shown in Fig.\ref{vortex_DS}(f)(i). All of the results are consistent with the above analysis. However, it is still puzzling why the there must be a VPT in the $H_{+i}\oplus H_{-i}$ subspace in doped superconducting DSs, which is obviously not the case in STIs\cite{Vishwanath_vortex}. This can be well understood if we solve the eigenstate problem at the Dirac point in the continuum limit by omitting the high-order terms. It turns out the zero mode corresponding to the VPT has the angular momentum\cite{supplementary}
\begin{eqnarray}\label{angular_matrix}
\hat{L}_z=-i\partial_\theta+\frac{1}{2}(\kappa-\Sigma_{33})\Pi_z+\frac{1}{2}\Pi_z,
\end{eqnarray}
where the first term originates from the rotation of the space, the second term $\frac{1}{2}(\kappa-\Sigma_{33})\Pi_z$ is the intrinsic angular momentum of the system which is determined by the basis of the Hamiltonian, while the third term stems from the quantum vortex line. Here, $\Pi_i$ are the Pauli matrices defined in the Nambu space. There is a key difference between the DSs and the STIs: $\kappa$ turns out to be 0 (1) for the STI (DSs), which results from the fact the band inversion in a 3D STI (DS) always occurs between two Kramers' doublets with the same (different) $J_z$.

\begin{figure}
\centerline{\includegraphics[width=0.45\textwidth]{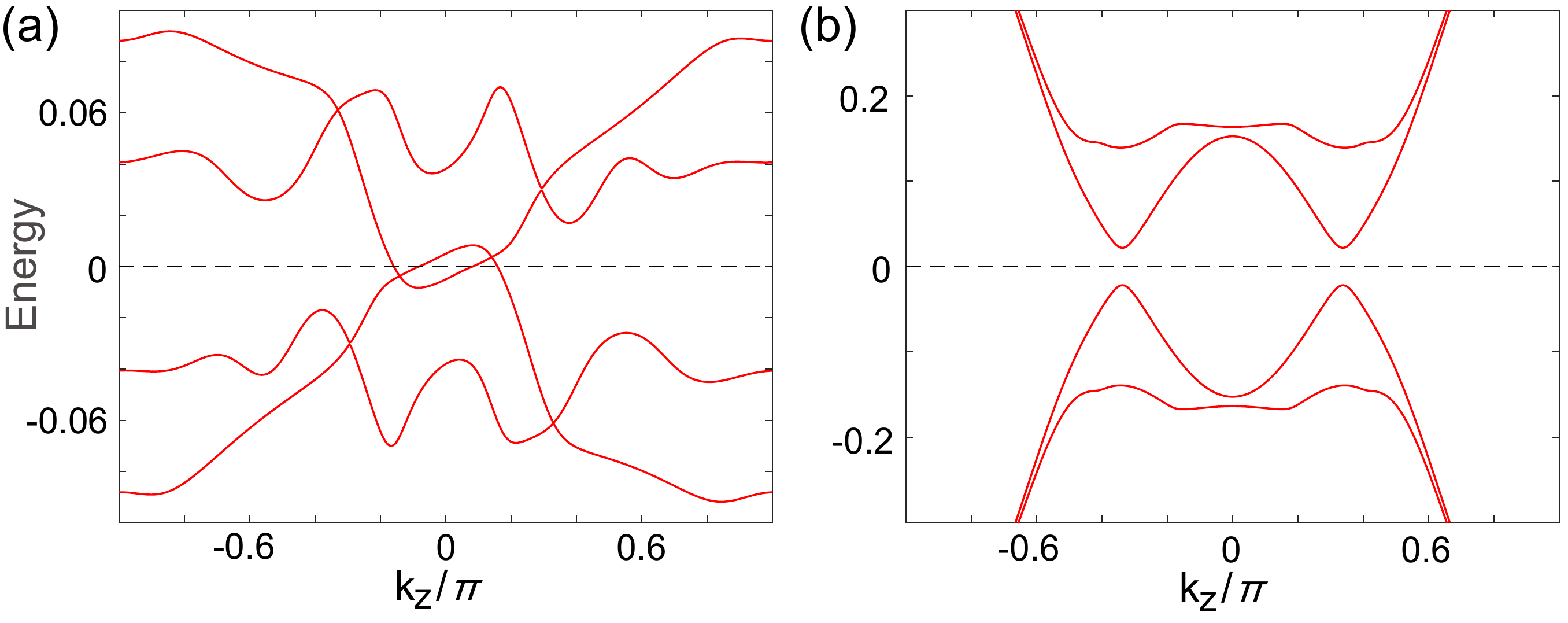}}
\caption{(color online) (a) shows the low-energy spectrum of the VBSs in the DS in $H_0^{(1)}$, when a IS breaking term $\frac{D}{2}\sin k_x(\Sigma_{01}+\Sigma_{31})+\frac{D}{2}\sin k_y(\Sigma_{02}+\Sigma_{32})$ is turned on. (b) corresponds to the case where there is a $C_{4z}$-rotational symmetry breaking perturbation $t_{sb}\sin k_z\Sigma_{11}$. Obviously, the IS breaking perturbation does not gap out the nodal vortex line phase, while $C_{4z}$-rotational symmetry breaking perturbation gaps out the nodal phase. We take $\{D, t_{sb}, \mu\}=\{2.0, 0.05, 0.24\}$ in the calculations.
\label{IS_break}}
\end{figure}

\emph{Effect of perturbation}.---We consider the effect of symmetry-breaking perturbations on the topological VPTs in this part. We first analyze the perturbations which break the IS but does not break the rotational-symmetry. This kind of perturbation lifts the band degeneracy in the spin degree of freedom, but it has no effect on the Dirac points\cite{Cr3As2}. As to the influence on the vortex bound states, since it does not break the rotational symmetry, the nodal vortex line phase is still robust for the sake that the nodal vortex line phase is protected by a finite energy gap, as shown in Fig.\ref{IS_break}(a). The above conclusions can be directly extended to other kinds of DSs. It is worth mentioning that for DSs protected by the $C_{3z}$-rotational symmetry, such kind of perturbations can break the Dirac point into a pair of triple points. Consequently, the nodal vortex line state can also be expected in triple-point semimetals\cite{triple1,triple2,triple3,triple4}.

Then we turn to perturbation which breaks the rotational symmetry. Such kind of perturbation can gap out the Dirac points, and the nontrivial DS will be broke into a STI. Correspondingly, the nodal vortex line will be broke into a full-gap topologically nontrivial phase, as shown in Fig.\ref{IS_break}(b), and MZMs emerge at the end of the vortex line. For the DS in $H_0^{(2)}$, the condition is complicated. However, there is a interesting case: if the $D_{4h}$ point group is broken into $D_{2h}$, the DS is broke into a TCI state with nonzero mirror Chern number protected by the mirror symmetry parallel to the $k_z$-axis\cite{topo_crystal}. In this case, the vortex line has a topologically nontrivial full-gap phase with multiple vortex end MZMs in a single vortex line\cite{Fang_vortex}.

\emph{Discussion and conclusion}.---In summary, we take the DSs protected by the $C_{4z}$-rotational symmetry as instances to study the VBSs in 3D superconducting DSs with TRS. We find that there is a robust quasi-1D nodal phase in the vortex line, if the DS can be doped to be $s$-wave superconducting. The nodal vortex line phase is protected by the $C_{4z}$-rotational symmetry and it stems from the symmetry protected Dirac points in the normal state bands. We also develop a topological index $(\nu; {\bf n})$ to characterize the nodal vortex line phase, where $\nu$ is the $Z_2$ topological invariant for a 0D class-$D$ system and ${\bf n}$ is the $Z$ topological invariant for a 0D class-$A$ system according to the AZ classification. Based on the classification, we find that the vortex end MZM can exist even in a gapless vortex line, and such a state can be realized in the trivial DS case. The influence of the symmetry breaking perturbations on the VBSs is also discussed, and we find that the vortex line can be broken into a topologically nontrivial full-gap state with vortex end MZMs for both the trivial and nontrivial DSs. The conclusions can be extended to general DSs with $C_{n}$-rotational symmetry ($n>2$), and the analysis in the text is also suitable for 3D TIs\cite{VPT_iron}.

Finally, we suggest that the nodal vortex line state can be realized in the nontrivial DSs Ir$_{1-x}$Pt$_x$Te$_2$\cite{IrTe2_SC1,IrTe2_SC2} and LiFeAs\cite{iron_TDS}, and the nodal vortex line state and vortex end MZMs can coexist in the trivial DS $\beta^\prime$-PtO$_2$\cite{PtO2}, if they can be $s$-wave superconducting when the chemical potential is doped near the Dirac points in their band structures. Especially, both the Dirac points in the band structure and the intrinsic superconductivity have been confirmed in Ir$_{1-x}$Pt$_x$Te$_2$ and LiFeAs.

S. S. Qin thanks H. M. Weng and X. X. Wu for helpful discussions. This work is supported by the Ministry of Science and Technology of China 973 program (Grant No. 2014CB921203, No. 2015CB921300, No. 2017YFA0303100), Ministry of Science and Technology of China (Grant No. 2016YFA0302400, 2016YFA0300600), National Science Foundation of China (Grant No. NSFC-11334012, No. NSFC-11674278), National Science Foundation of China (Grant No. 11674370), and the Strategic Priority Research Program of CAS (Grant No. XDB07000000).

\emph{Note added}: During the preparation of this manuscript, we notice that a preprint\cite{nodal_vortex} also discuss the nodal vortex line phase in iron-based SCs when the chemical potential is near the Dirac point, which is consistent with our results.


%
%


\appendix

\begin{widetext}

\section{Topological invariant for the $1D$ class-$D$ superconductors}

As mentioned in the main text, if there is a quantum vortex in a 2D spinful SC, the whole system can be viewed as a quasi-0D system belonging to the class-$D$ of the Altland-Zirnbauer (AZ) classification. Such a system has a $Z_2^{0D}$ topological invariant which characterizes its fermion parity. To calculate the $Z_2^{0D}$ topological invariant, we can first rewrite the Hamiltonian $H_{sc}({\bf k})$ in the Majorana basis
\begin{eqnarray}\label{Majorana_basis}
\gamma_1^\dag=\frac{1}{\sqrt{2}}[c_\uparrow^\dag+c_\uparrow],
\gamma_2^\dag=\frac{1}{\sqrt{2}}[c_\downarrow^\dag+c_\downarrow], \nonumber\\
\gamma_3^\dag=\frac{i}{\sqrt{2}}[c_\uparrow^\dag-c_\uparrow],
\gamma_4^\dag=\frac{i}{\sqrt{2}}[c_\downarrow^\dag-c_\downarrow],
\end{eqnarray}
where $\psi^\dag=(c_\uparrow^\dag, c_\downarrow^\dag, c_\uparrow, c_\downarrow)$ is the traditional basis for the BdG Hamiltonian. In the Majorana basis, the Hamiltonian can be expressed as $H_{sc}({\bf k})=i\gamma^\dag A\gamma$, where $A$ is a real antisymmetric matrix and its Pfaffian is well defined, and the $Z_2^{0D}$ topological index has the form
\begin{eqnarray}\label{0D_topo_invariant}
\nu=sgn(Pf(A)),
\end{eqnarray}
where $sgn()$ is the signum and $Pf()$ is the Pfaffian.

a 3D spinful SC with a vortex line belongs to the 1D class-$D$ of the AZ classification. For such a 1D system, if the quasi-particle spectrum is full-gap, it has a $Z_2^{1D}$ topological classification, which is also characterized by the fermion parity of the system. As pointed out, the $Z_2^{1D}$ topological invariant can be calculated as follows
\begin{eqnarray}\label{1D_topo_invariant}
\nu^{1D}=sgn(Pf(A(k=0)))\cdot sgn(Pf(A(k=\pi))),
\end{eqnarray}
similarly, $A$ is the Hamiltonian in the Majorana basis here. Obviously, the above equation can also be expressed as
\begin{eqnarray}\label{1D_topo_invariant}
\nu^{1D}=\nu(k=0)\cdot \nu(k=\pi).
\end{eqnarray}


\section{Construction of the $C_{4z}$-rotational symmetry for the vortex line system}

In this section, we show how to derive the matrix form of $C_{4z}^{vortex}$ in the main text. We consider a geometry which has finite size in the $xy$-plane and respects the $C_{4z}$-rotational symmetry. If there is no vortex line in the superconducting DS, we can straightforwardly write down the matrix form of the $C_{4z}$-rotational symmetry operator in the basis $\psi^\dag=(c_{1/2,k_z}^\dag, c_{3/2,k_z}^\dag,c_{-1/2,k_z}^\dag, c_{-3/2,k_z}^\dag, c_{1/2,-k_z}, c_{3/2,-k_z},c_{-1/2,-k_z}, c_{-3/2,-k_z})$ as follows
\begin{eqnarray}\label{C4_BdG}
C_{4z}^{sc}=\delta_{ij}\left(
            \begin{array}{cc}
            C_{4z}                       &                0                  \\
            0                            &            C_{4z}^\ast            \\
            \end{array}
          \right),
\end{eqnarray}
where $C_{4z}$ is the matrix form of the $C_{4z}$-rotational operator in the normal state in the main text, and $i$ and $j$ are the site index in the $xy$-plane which are related by the $C_{4z}$-rotational symmetry. If a single quantum vortex line is added, the creation and annihilation operators gain a $U(1)$ gauge field. Correspondingly, the matrix form of the $C_{4z}$-rotational operator is transformed into
\begin{eqnarray}\label{C4_BdG_vortex}
C_{4z}^{vortex}=\delta_{ij}\left(
            \begin{array}{cc}
            C_{4z}e^{i\pi/4}             &                      0                      \\
            0                            &            C_{4z}^\ast e^{-i\pi/4}          \\
            \end{array}
          \right).
\end{eqnarray}
Obviously, without vortex the $C_{4z}$-rotational symmetry takes the eigenvalues $e^{i(2n+1)\pi/4}$ ($n$ is a integer) while its eigenvalues are $e^{in\pi/2}$ after the quantum vortex line is added.

\begin{table}[bt]
\caption{\label{Z classify} In the weak pairing limit, the $Z^{0D}$ topological number $N$ in each $C_{4z}^{vortex}$ subspace at the time reversal invariant momenta for the trivial and nontrivial DSs in the main text is summarized in the table. In the calculations, all of the parameters are all the same as that in the main text.}
\begin{ruledtabular}
\begin{tabular}{c|cccc}
                                  & $N_{+i}$  & $N_{-1}$  & $N_{-i}$ & $N_{+1}$ \\
\colrule
trivial DS at $k_z=0$             & 577       & 576       & 575      & 576 \\
trivial DS at $k_z=\pi$           & 576       & 576       & 576      & 576 \\
\hline
nontrivial DS $k_z=0$             & 577       & 576       & 575      & 576 \\
nontrivial DS $k_z=\pi$           & 576       & 576       & 576      & 576 \\
\end{tabular}
\end{ruledtabular}
\end{table}

Based on the above constructed matrix form of the $C_{4z}$-rotational symmetry, we have calculated the $Z^{0D}$ topological index in each $C_{4z}^{vortex}$ subspace numerically and the results are summarized in Table.\ref{Z classify}.

\section{Berry phase on the FSs in the $k_z=0$ plane}

In this section, we use the continuum models to analyze the VPTs at the TRI momenta for different DSs with linear Dirac points, based on the Berry phase criterion. We take all kinds of 3D DSs which have TRS, IS and uniaxial rotational symmetry into consideration.

\subsection{$C_4$-case}

The nontrivial DSs protected by the $C_4$ rotational symmetry have the following general Hamiltonian
\begin{eqnarray}\label{TDSC_4}
H_{4T}=\left(
            \begin{array}{cccc}
            m({\bf k})                       &                Ak_+                &                0                     &  \alpha k_zk_-^2+\beta k_zk_+^2   \\
            Ak_-                             &            -m({\bf k})             &    \alpha k_zk_-^2+\beta k_zk_+^2    &                 0                 \\
            0                                &   \alpha k_zk_+^2+\beta k_zk_-^2   &            m({\bf k})                &               -Ak_-               \\
            \alpha k_zk_+^2+\beta k_zk_-^2   &                 0                  &              -Ak_+                   &              -m({\bf k})          \\
            \end{array}
          \right),
\end{eqnarray}
where $m({\bf k})=m-t_3k_z^2+t(k_x^2+k_y^2)$, $k_\pm=k_x\pm ik_y$. The basis for the above Hamiltonian is $(|\frac{1}{2},+\rangle, |\frac{3}{2},-\rangle, |-\frac{1}{2},+\rangle, |-\frac{3}{2},-\rangle)$, where $J_z$ is the angular momentum defined according to the $C_4$ rotational axis and $s$ is its parity in $|J_z,s\rangle$. For the $C_4$-case, since there is also the IS, the system must also have a mirror symmetry $I(C_4)^2$, which reflects $z$ to $-z$. As a result, the DS can be decoupled into different mirror subspaces and the $SU(2)$ Berry phase on the two-fold degenerate Fermi surfaces (FSs) can be reduced into the $U(1)$ Berry phase in different subspaces in the $k_z=0$ plane, which is also true for the case which has the $C_6$ rotational symmetry. In the following calculations, we only consider the $U(1)$ Berry phase in the subspace with mirror eigenvalue $+i$ if there are no special instructions. For the DS depicted by $H_{4T}$, we consider the reduced Hamiltonian, $H_{4T}^{re}=m({\bf k})\sigma_z+Ak_x\sigma_x-Ak_y\sigma_y$, which can be solved easily
\begin{eqnarray}\label{eigenvalue_TDSC_4}
\pm E=\pm\sqrt{A^2k^2+m^2({\bf k})}, \quad
|+E\rangle=\frac{1}{a}\left(
            \begin{array}{c}
              Ake^{i\theta} \\
              E-m({\bf k})     \\
            \end{array}
          \right),
|-E\rangle=\frac{1}{b}\left(
            \begin{array}{c}
              -Ake^{i\theta} \\
              E+m({\bf k})      \\
            \end{array}
          \right),
\end{eqnarray}
where we have write ${\bf k}$ in the polar coordinate system, $k^2=k_+k_-$, and $a$ and $b$ are the corresponding normalization coefficient. The Berry phase on the FS can be figured out
\begin{eqnarray}\label{Berry_TDSC_4}
\varphi&=&i\oint_{FS}d{\bf k}\langle+E|\nabla_{{\bf k}}|+E\rangle=\int d\theta k\cdot \langle+E|\frac{\partial_\theta}{k}|+E\rangle
=-\pi(1+\frac{m({\bf k})}{\sqrt{m^2({\bf k})+A^2k^2}}),
\end{eqnarray}
Straightforwardly, we can conclude that: if $m/t>0$, $\varphi$ can never reach $\pi$ (mod $2\pi$); if $m/t=0$, $\varphi$ equals to $\pi$ at the limit $k=0$; if $m/t<0$, $\varphi$ vanishes at the limit $k=0$, and it reaches $\pi$ when $m+tk^2$ is zero. Obviously, only the case corresponding to a band inversion at the $\Gamma$ point robustly has a chemical potential, at which the Berry phase on the FS can reach $\pi$.

\begin{figure}
\centerline{\includegraphics[width=0.5\textwidth]{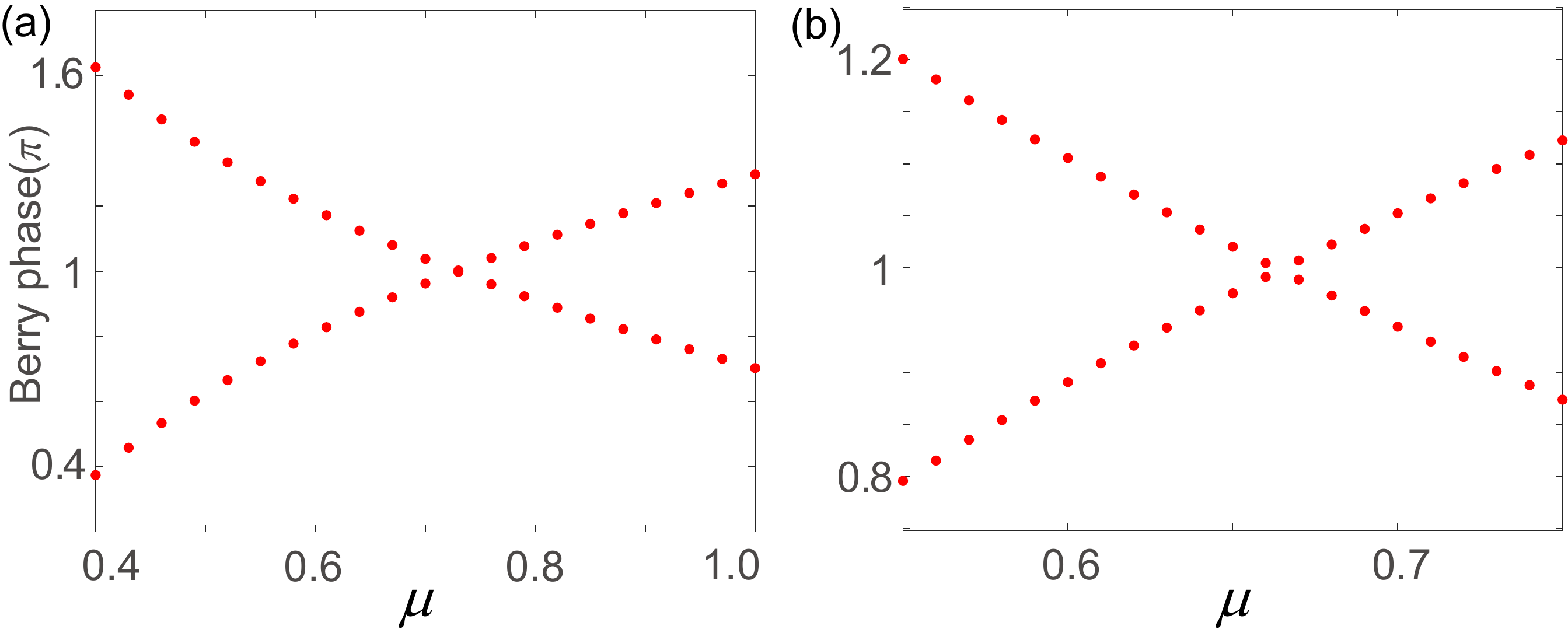}}
\caption{(color online) The $SU(2)$ Berry phase on the FSs as a function of the chemical potential $\mu$ in the $k_z=0$ plane for the trivial DS in $H_{4N}$. Obviously, it reaches $\pm\pi$ at $\mu_c^1=0.353$ and $\mu_c^2=0.985$. In the calculations, we take the parameters as $\{A, \alpha, \beta\}=\{1.0, 1.0, 0\}$ and set $m({\bf k})=m-t_3k_z^2-tk^2-t_4k^4$, where $\{m, t, t_3, t_4\}=\{0.5, 1.0, 0.5, 0.2\}$.
\label{berry_quadratic}}
\end{figure}

The Hamiltonian describing the trivial DSs protected by the $C_4$ rotational symmetry have the form
\begin{eqnarray}\label{NDSC_4}
H_{4N}=\left(
            \begin{array}{cccc}
            m({\bf k})                       &              Ak_zk_+              &                0                &    \alpha k_-^2+\beta k_+^2   \\
            Ak_zk_-                          &            -m({\bf k})            &    -\alpha k_-^2-\beta k_+^2    &                0              \\
            0                                &     -\alpha k_+^2-\beta k_-^2     &            m({\bf k})           &              Ak_zk_-          \\
            \alpha k_+^2+\beta k_-^2         &                 0                 &              Ak_zk_+            &             -m({\bf k})       \\
            \end{array}
          \right),
\end{eqnarray}
in which the basis is $(|\frac{1}{2},+\rangle, |\frac{3}{2},+\rangle, |-\frac{1}{2},+\rangle, |-\frac{3}{2},+\rangle)$. Similarly, we can decouple $H_{4N}$ in the $k_z=0$ plane, $H_{4N}^{re}=m({\bf k})\sigma_z+\alpha(k_x^2-k_y^2)\sigma_x+2\alpha k_xk_y\sigma_y$, where we have set $\beta=0$ to make the system has continuum rotational symmetry, and solve it
\begin{eqnarray}\label{eigenvalue_NDSC_4}
\pm E=\pm\sqrt{\alpha^2k^4+m^2({\bf k})}, \quad
|+E\rangle=\frac{1}{a}\left(
            \begin{array}{c}
              \alpha k^2e^{-2i\theta} \\
              E-m({\bf k})     \\
            \end{array}
          \right),
|-E\rangle=\frac{1}{b}\left(
            \begin{array}{c}
              -\alpha k^2e^{-2i\theta} \\
              E+m({\bf k})      \\
            \end{array}
          \right).
\end{eqnarray}
The Berry phase on the FS can also be calculated
\begin{eqnarray}\label{Berry_NDSC_4}
\varphi&=&i\oint_{FS}d{\bf k}\langle+E|\nabla_{{\bf k}}|+E\rangle=\int d\theta k\cdot \langle+E|\frac{\partial_\theta}{k}|+E\rangle
=2\pi(1+\frac{m({\bf k})}{\sqrt{m^2({\bf k})+\alpha^2k^4}}).
\end{eqnarray}
From Eq.\ref{Berry_NDSC_4}, it can be worded out easily that when $3m^2({\bf k})=\alpha^2k^4$, the Berry phase on the FS is $\pi$. It should be mentioned that, to make sure that the vortex line is topologically trivial when the chemical potential is infinite large, we need to consider high-order terms in $m({\bf k})$, for instance $m({\bf k})=m+tk^2+\epsilon k^4$. To make a band inversion at the $\Gamma$ point, we set $m/\epsilon <0$. It is easy to figure out that, the above equation have $2+2n$ ($n\geq 0$ and is a integer) solutions, namely the corresponding vortex line has even number of VPTs at $k_z=0$, which is consistent with the results in the main text. Fig.\ref{berry_quadratic} shows the $SU(2)$ Berry phase on the FSs for $H_{4N}$ in the $k_z=0$ plane as a function of the chemical potential. Obviously, it reaches $\pm\pi$ at $\mu_c^1=0.985$ and $\mu_c^2=0.353$.


\subsection{$C_3$-case}

The nontrivial DSs protected by the $C_3$ rotational symmetry have the following general Hamiltonian
\begin{eqnarray}\label{TDSC_3}
H_{3T}=\left(
            \begin{array}{cccc}
            m({\bf k})   &       Ak_+        &        0         &  \alpha k_+   \\
            Ak_-         &     -m({\bf k})   &    \alpha k_+    &      0        \\
            0            &  \alpha k_-       &     m({\bf k})   &    -Ak_-      \\
            \alpha k_-   &        0          &      -Ak_+       &  -m({\bf k})  \\
            \end{array}
          \right),
\end{eqnarray}
where the basis is $(|\frac{1}{2},+\rangle, |\frac{3}{2},-\rangle, |-\frac{1}{2},+\rangle, |-\frac{3}{2},-\rangle)$. For the $C_3$-case, we can not decouple the Hamiltonian in the $k_z=0$ plane since there is no mirror symmetry perpendicular to the $z$-axis. However, we can generally calculate the $SU(2)$ Berry phase on the FSs. We first solve $H_{3T}$ and achieve its eigenfunctions for the conduction bands as follows
\begin{eqnarray}\label{eigenvalue_TDSC_3}
\pm E=\pm\sqrt{A^2k^2+\alpha^2k^2+m^2({\bf k})}, \quad
|+E,1\rangle=\frac{1}{a}\left(
            \begin{array}{c}
              E+m({\bf k})           \\
              Ake^{-i\theta}         \\
              0                      \\
              \alpha ke^{-i\theta}   \\
            \end{array}
          \right),
|+E,2\rangle=\frac{1}{b}\left(
            \begin{array}{c}
              0                        \\
              -\alpha ke^{i\theta}     \\
              -E-m({\bf k})            \\
              Ake^{i\theta}            \\
            \end{array}
          \right).
\end{eqnarray}
We can get the $SU(2)$ Berry phase on the FSs as
\begin{eqnarray}\label{Berry_TDSC_3}
\varphi&=&i\oint_{FS}d{\bf k}\langle +E,i|\nabla_{{\bf k}}|+E,j\rangle=\int d\theta k\cdot \langle +E,i|\frac{\partial_\theta}{k}|+E,j\rangle    \nonumber\\
&=&\frac{\pi k^2}{E(E+m({\bf k}))}\left(
                                  \begin{array}{cc}
                                  \alpha^2+A^2     &      0                  \\
                                  0                &      -\alpha^2-A^2      \\
                                  \end{array}
                                  \right)
=\pi\left(
          \begin{array}{cc}
          1-\frac{m({\bf k})}{E}     &       0                           \\
          0                          &      -1+\frac{m({\bf k})}{E}      \\
          \end{array}
          \right),
\end{eqnarray}
in which $i, j$ are the indexes for the eigenfunctions of the conduction bands.

The corresponding trivial DSs protected by the $C_3$ rotational symmetry have the Hamiltonian
\begin{eqnarray}\label{NDSC_3}
H_{3N}=\left(
            \begin{array}{cccc}
            m({\bf k})                  &         Ak_zk_++Bk_-^2           &                0                 &    \alpha k_zk_++\beta k_-^2    \\
            Ak_zk_-+Bk_+^2              &          -m({\bf k})             &    -\alpha k_zk_+-\beta k_-^2    &                0                \\
            0                           &    -\alpha k_zk_--\beta k_+^2    &             m({\bf k})           &          Ak_zk_-+Bk_+^2         \\
            \alpha k_zk_-+\beta k_+^2   &               0                  &           Ak_zk_++Bk_-^2         &            -m({\bf k})          \\
            \end{array}
          \right),
\end{eqnarray}
in which the basis is $(|\frac{1}{2},+\rangle, |\frac{3}{2},+\rangle, |-\frac{1}{2},+\rangle, |-\frac{3}{2},+\rangle)$. Similarly, we can solve the above Hamiltonian
\begin{eqnarray}\label{eigenvalue_NDSC_3}
\pm E=\pm\sqrt{B^2k^4+\beta^2k^4+m^2({\bf k})}, \quad
|+E,1\rangle=\frac{1}{a}\left(
            \begin{array}{c}
              E+m({\bf k})             \\
              Bk^2e^{2i\theta}         \\
              0                        \\
              \beta k^2e^{2i\theta}    \\
            \end{array}
          \right),
|+E,2\rangle=\frac{1}{b}\left(
            \begin{array}{c}
              0                           \\
              -\beta k^2e^{-2i\theta}     \\
              -E-m({\bf k})               \\
              Bk^2e^{-2i\theta}           \\
            \end{array}
          \right).
\end{eqnarray}
The $SU(2)$ Berry phase on the FSs are
\begin{eqnarray}\label{Berry_NDSC_3}
\varphi&=&i\oint_{FS}d{\bf k}\langle +E,i|\nabla_{{\bf k}}|+E,j\rangle=\int d\theta k\cdot \langle +E,i|\frac{\partial_\theta}{k}|+E,j\rangle    \nonumber\\
&=&\frac{2\pi k^4}{E(E+m({\bf k}))}\left(
                                  \begin{array}{cc}
                                  \beta^2+B^2     &      0                  \\
                                  0               &      -\beta^2-B^2       \\
                                  \end{array}
                                  \right)
=2\pi\left(
          \begin{array}{cc}
          -1+\frac{m({\bf k})}{E}     &       0                           \\
          0                           &       1-\frac{m({\bf k})}{E}      \\
          \end{array}
          \right).
\end{eqnarray}

Compare Eq.\ref{Berry_TDSC_3} and \ref{Berry_NDSC_3} with Eq.\ref{Berry_TDSC_4} and \ref{Berry_NDSC_4}, we can find that the results are quite similar. Therefore, we can conclude that the DSs protected by the $C_3$ rotational symmetry have similar VPTs with the $C_4$ case when $k_z=0$. It should be mentioned that, we can calculate the $SU(2)$ Berry phase analytically here, since the off-diagonal terms of the non-Abelian Berry connection are zero at each $k$ on the integral path, namely $\langle i|\nabla_{{\bf k}}|j\rangle=0$ for $i\neq j$.

\subsection{$C_6$-case}

The DSs with linear Dirac points protected by the $C_6$ rotational symmetry can be classified into two classes according to its baiss.

\subsubsection{case-\uppercase\expandafter{\romannumeral1}}
In the first case, the nontrivial DSs have the following general Hamiltonian
\begin{eqnarray}\label{TDSC_61}
H_{6T}^{\uppercase\expandafter{\romannumeral1}}=\left(
            \begin{array}{cccc}
            m({\bf k})         &           Ak_+           &           0            &    \alpha k_zk_-^2    \\
            Ak_-               &       -m({\bf k})        &    \alpha k_zk_-^2     &           0           \\
            0                  &     \alpha k_zk_+^2      &       m({\bf k})       &         -Ak_-         \\
            \alpha k_zk_+^2    &            0             &         -Ak_+          &       -m({\bf k})     \\
            \end{array}
          \right),
\end{eqnarray}
where the basis is $(|\frac{1}{2},+\rangle, |\frac{3}{2},-\rangle, |-\frac{1}{2},+\rangle, |-\frac{3}{2},-\rangle)$.

The corresponding trivial DSs have the form
\begin{eqnarray}\label{NDSC_61}
H_{6N}^{\uppercase\expandafter{\romannumeral1}}=\left(
            \begin{array}{cccc}
            m({\bf k})        &        Ak_zk_+        &          0          &    \alpha k_-^2     \\
            Ak_zk_-           &      -m({\bf k})      &   -\alpha k_-^2     &          0          \\
            0                 &     -\alpha k_+^2     &      m({\bf k})     &        Ak_zk_-      \\
            \alpha k_+^2      &          0            &       Ak_zk_+       &      -m({\bf k})    \\
            \end{array}
          \right),
\end{eqnarray}
in which the basis is $(|\frac{1}{2},+\rangle, |\frac{3}{2},+\rangle, |-\frac{1}{2},+\rangle, |-\frac{3}{2},+\rangle)$.

\subsubsection{case-\uppercase\expandafter{\romannumeral2}}

In the second case, the nontrivial DSs have the following general Hamiltonian
\begin{eqnarray}\label{TDSC_62}
H_{6T}^{\uppercase\expandafter{\romannumeral2}}=\left(
            \begin{array}{cccc}
            m({\bf k})         &           Ak_+           &           0            &    \alpha k_zk_+^2    \\
            Ak_-               &       -m({\bf k})        &    \alpha k_zk_+^2     &           0           \\
            0                  &     \alpha k_zk_-^2      &       m({\bf k})       &         -Ak_-         \\
            \alpha k_zk_-^2    &            0             &         -Ak_+          &       -m({\bf k})     \\
            \end{array}
          \right),
\end{eqnarray}
where the basis is $(|\frac{3}{2},+\rangle, |\frac{5}{2},-\rangle, |-\frac{3}{2},+\rangle, |-\frac{5}{2},-\rangle)$.

The corresponding trivial DSs have the form
\begin{eqnarray}\label{NDSC_62}
H_{6N}^{\uppercase\expandafter{\romannumeral2}}=\left(
            \begin{array}{cccc}
            m({\bf k})        &        Ak_zk_+        &          0          &    \alpha k_+^2     \\
            Ak_zk_-           &      -m({\bf k})      &   -\alpha k_+^2     &          0          \\
            0                 &     -\alpha k_-^2     &      m({\bf k})     &        Ak_zk_-      \\
            \alpha k_-^2      &          0            &       Ak_zk_+       &      -m({\bf k})    \\
            \end{array}
          \right),
\end{eqnarray}
in which the basis is $(|\frac{3}{2},+\rangle, |\frac{5}{2},+\rangle, |-\frac{3}{2},+\rangle, |-\frac{5}{2},+\rangle)$.

It can be notice that, for the DSs protected by the $C_6$ rotational symmetry, all the Hamiltonians are equivalent to $H_{4T}$ or $H_{4N}$ when $k_z=0$, which means that these DSs have similar VPTs at $k_z=0$ with the $C_4$-case.

Remarkably, in the continuum limit when and only when a system has band inversion, its corresponding superconducting vortex line can have VPTs. This is attributed to the fact that, its Berry phase on the FS is 0 at the limit $k\rightarrow 0$ while $2n\pi$ when $k\rightarrow\infty$ (we have assume that $m<0$ and the coefficient of the highest $k$ order $\epsilon>0$). If there is no band inversion, namely $m>0$, the Berry phase on the FS is $2n\pi$ both at the limit $k\rightarrow 0$ and $k\rightarrow\infty$. Therefore, there may be even number of VPT points accidentally which are sensitive to the details of the coefficients in the Hamiltonian.

\section{Comparison with the 3D STI Case}

In this section, we take the DSs and the STIs in the $C_{4z}$-rotational invariant case to explain why the superconducting vortex line in a 3D DS must have a nodal phase while it has a topologically nontrivial full-gap phase in a 3D STI. We can write down a general continuum model for a 3D STI as follows
\begin{eqnarray}\label{STIC_4}
H_{TI}=\left(
            \begin{array}{cccc}
            m({\bf k})                       &             \alpha k_z             &                0                     &               Ak_-                \\
            \alpha k_z                       &            -m({\bf k})             &               Ak_-                   &                 0                 \\
            0                                &                Ak_+                &            m({\bf k})                &           -\alpha k_z             \\
            Ak_+                             &                 0                  &            -\alpha k_z               &            -m({\bf k})            \\
            \end{array}
          \right),
\end{eqnarray}
where the basis is $(|\frac{1}{2},+\rangle, |\frac{1}{2},-\rangle, |-\frac{1}{2},+\rangle, |-\frac{1}{2},-\rangle)$.

\subsection{at $k_z=0$}

We compare the DSs with the STIs in two aspects: (i) the Berry phase on the FS, and (ii) the spectrum of the bound states in the superconducting vortex line.

\subsubsection{Berry phase on the FS}

Since $H_{TI}$ has band inversion at the $\Gamma$ point, we first calculate the Berry phase on the FSs in the $k_z=0$ plane. Similar to the DS case, we can decouple the Hamiltonian according to the mirror symmetry perpendicular to the $k_z$-axis $M_z$, and consider the reduced Hamiltonian in the mirror $+i$ subspace, $H_{TI}^{re}=m({\bf k})\sigma_z+Ak_x\sigma_x+Ak_y\sigma_y$, which can be solved as
\begin{eqnarray}\label{eigenvalue_STIC_4}
\pm E=\pm\sqrt{A^2k^2+m^2({\bf k})}, \quad
|+E\rangle=\frac{1}{a}\left(
            \begin{array}{c}
              Ake^{-i\theta} \\
              E-m({\bf k})     \\
            \end{array}
          \right),
|-E\rangle=\frac{1}{b}\left(
            \begin{array}{c}
              -Ake^{-i\theta} \\
              E+m({\bf k})      \\
            \end{array}
          \right).
\end{eqnarray}
The Berry phase on the FS is
\begin{eqnarray}\label{Berry_STIC_4}
\varphi&=&i\oint_{FS}d{\bf k}\langle+E|\nabla_{{\bf k}}|+E\rangle=\int d\theta k\cdot \langle+E|\frac{\partial_\theta}{k}|+E\rangle
=\pi(1+\frac{m({\bf k})}{\sqrt{m^2({\bf k})+A^2k^2}}).
\end{eqnarray}
Obviously, it has similar results with the DS case and we can not tell the difference between the VPTs in the two kinds of systems.

\subsubsection{Bound states in the vortex line}

As analyzed in the main text, the VPT occurs at the topological phase transition points of the normal state band structures. In the following, we focus on the vortex bound states at the topological phase transition point of the normal state band structures when $\mu=0$ for the superconducting DS and STI.

Since the mirror symmetry $M_z$ is always true no matter there is a vortex line or not, we can decouple the superconducting Hamiltonian according to $M_z$. In the following, we only consider the mirror $+i$ subspace, for the sake that the particle-hole symmetry (PHS) $P$ transforms the mirror $+i$ subspace into the $-i$ subspace. We first consider the superconducting Hamiltonian without vortex lines, and find that if we choose the gauge properly, both the superconducting DS and STI have the same Hamiltonian in the mirror $+i$ subspace
\begin{eqnarray}\label{Dirac_equation}
H({\bf k})=\left(
            \begin{array}{cccc}
            0                   &           0            &         \Delta            &         Ak_-           \\
            0                   &           0            &          Ak_+             &       -\Delta          \\
            \Delta              &          Ak_-          &            0              &           0            \\
            Ak_+                &        -\Delta         &            0              &           0            \\
            \end{array}
          \right),
\end{eqnarray}
where the basis is $\Psi_{TI}^\dag({\bf k})=(c_{-1/2,+}(-{\bf k}),c_{-1/2,-}^\dag({\bf k}),c_{1/2,+}^\dag({\bf k}),c_{1/2,-}(-{\bf k}))$ for the STI and $\Psi_{DS}^\dag({\bf k})=(c_{3/2,-}^\dag({\bf k}),c_{-1/2,+}(-{\bf k}),-c_{-3/2,-}(-{\bf k}),c_{1/2,+}^\dag({\bf k}))$ for the DS. Notice that Eq.\ref{Dirac_equation} is exactly the form of the 3D Dirac equation. If a infinite small quantum vortex line is added (we set the vortex line along the $+z$-direction for the STI and $-z$-direction for the DS, the two systems still has the same Hamiltonian under the gauge in Eq.\ref{Dirac_equation}), we can solve the system in the real space
\begin{eqnarray}\label{Dirac_equation_vortex}
H({\bf k})=\left(
            \begin{array}{cccc}
            0                 &              0              &           \Delta e^{-i\theta}          &              Ae^{-i\theta}(-i\partial_r-\frac{\partial_\theta}{r})   \\
            0                 &              0              &            Ae^{i\theta}(-i\partial_r+\frac{\partial_\theta}{r})                &       -\Delta e^{i\theta}    \\
            \Delta e^{i\theta}       &            Ae^{-i\theta}(-i\partial_r-\frac{\partial_\theta}{r})             &            0           &             0                \\
            Ae^{i\theta}(-i\partial_r+\frac{\partial_\theta}{r})       &       -\Delta e^{-i\theta}        &               0                 &              0               \\
            \end{array}
          \right),
\end{eqnarray}
and it is easy to figure out that the above Hamiltonian has a zero mode, whose eigenfuction is
\begin{eqnarray}\label{zero mode}
\varphi_0(r)=e^{-\int_0^r|\Delta(r)/A|dr^\prime}(0,0,sgn(A\Delta)i,-sgn(A\Delta)i)^T.
\end{eqnarray}

With the gauge in Eq.\ref{Dirac_equation}, the VPT points for both the DS and STI cases share the same eigenfuction in Eq.\ref{zero mode}. Therefore, it seems that the vortex line in a doped superconducting STI and DS should have similar topological properties, which conflicts with the numerical results apparently. To solve the contradiction, we need to consider the physical meaning of Eq.\ref{Dirac_equation}. In the field of particle physics, it is well known that Eq.\ref{Dirac_equation} describes a Dirac fermion and the vacuum is always isotropic. In condensed matter physics, Eq.\ref{Dirac_equation} describes a Dirac fermion like quasi-particle, however, the "vacuum" here is not always isotropic. This can be well illustrated by constructing the angular momentum operator of Eq.\ref{Dirac_equation_vortex}
\begin{eqnarray}\label{angular_matrix}
\hat{L}_z=-i\partial_\theta+
            \frac{1}{2}\left(
            \begin{array}{cccc}
            1        &     0     &     0      &      0        \\
            0        &    -1     &     0      &      0        \\
            0        &     0     &     1      &      0        \\
            0        &     0     &     0      &     -1        \\
            \end{array}
          \right)+
          \frac{1}{2}\left(
            \begin{array}{cccc}
            1        &     0     &     0      &      0        \\
            0        &    -1     &     0      &      0        \\
            0        &     0     &    -1      &      0        \\
            0        &     0     &     0      &      1        \\
            \end{array}
          \right)+\kappa I_4,
\end{eqnarray}
where $I_4$ is the identity matrix, $\kappa$ is the angular momentum of the "vacuum", and $\hat{L}_z-\kappa I_4$ is the angular momentum of the Dirac fermion. Obviously, the zero mode in Eq.\ref{zero mode} has angular momentum 0 if the "vacuum" is not take into consideration. However, by analyzing the basis of Eq.\ref{Dirac_equation}, it is easy to find that the "vacuum" in the STI (DS) case carries an angular momentum of 0 (1), which means that the $L_z$ of the zero mode is 0 (1) for the STI (DS). Recalling that the PHS transforms the mirror subspace with eigenvalue $+i$ into the $-i$ one, we immediately know that there is a zero mode with angular momentum 0 (1) in the mirror $-i$ subspace for the STI (DS). Based on the above results, it can be straightforwardly concluded that the the zeros modes must be in the $H_{+1}$ ($H_{+i}\oplus H_{-i}$) subspace for the STI (DS) case if the continuum rotational symmetry is broke into the $C_{4z}$ one, which is consistent with the numerical results in the main text.

\begin{figure}
\centerline{\includegraphics[width=0.5\textwidth]{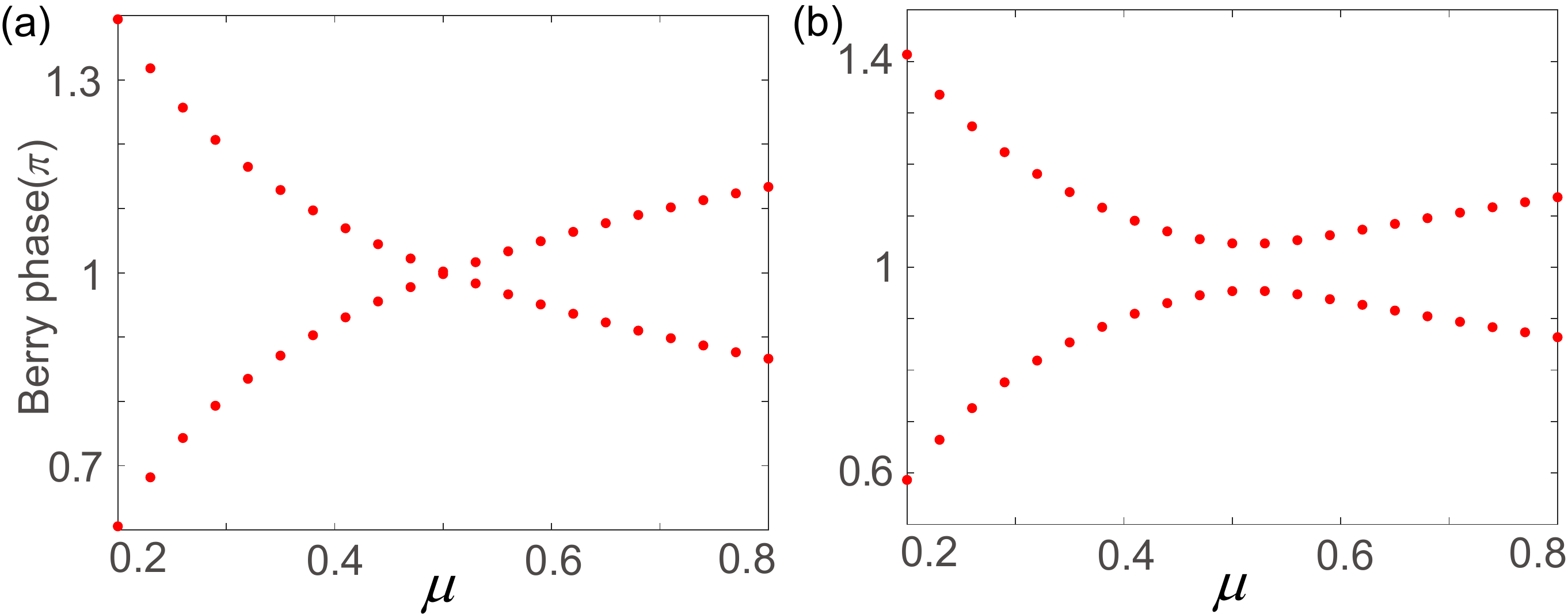}}
\caption{(color online) The $SU(2)$ Berry phase on the FSs for the $Z_2^{DS}$ nontrivial DS case in the main text as a function of the chemical potential $\mu$ at $k_z=0.3\pi$, is shown in (a). Obviously, it reaches $\pm\pi$ at $\mu_c=0.50$. When a rotational symmetry breaking term $t_{sb}\sin k_z\Sigma_{11}$ is added, the DS becomes a STI. In this case, the $SU(2)$ Berry phase on the FSs as a function of $\mu$ at $k_z=0.3\pi$, is shown in (b). Clearly, the $SU(2)$ Berry phase can never reach $\pm\pi$ in (b). We take $t_{sb}=0.03$ in the calculation, and all the other parameters are the same with that in the main text.
\label{Berry k no 0}}
\end{figure}

\subsection{at $k_z \neq 0$}

Then, we turn to the $k_z \neq 0$ case, and we consider the Berry phase on the FSs and the vortex bound states similarly.

\subsubsection{Berry phase on the FS}

Generally, there is a key difference between the STIs and DSs: a 3D STI merely has band inversions at the time reversal invariant points in the Brillouin zone (BZ), while a DS has band inversions at each $k$-point in between the two Dirac points. Specifically, the STI in Eq.\ref{STIC_4} merely has band inversion at (0,0,0), while the DS in Eq.\ref{TDSC_4} has band inversion at each (0, 0, $k_z$) for $|k_z|<\sqrt{\frac{m}{t_3}}$. It is easy to figure out that, the term $\alpha k_z\Sigma_{13}$ acts as a mass term which prevent the band inversions at $k_z\neq 0$, while such a term is forbidden by symmetries in a DS. This results in that, the $SU(2)$ Berry phase on the FSs in a given $k_z\neq 0$ plane can never reaches $\pi$ for STIs, while it can be realized at some critical chemical potential in each $|k_z|<\sqrt{\frac{m}{t_3}}$ plane for the DSs. We have checked this numerically based on the $Z_2^{DS}$ nontrivial DS in the main text, as shown in Fig.\ref{Berry k no 0}.

\subsubsection{Bound states in the vortex line}

Since the mirror symmetry $M_z$ no longer exists at $k_z\neq 0$, it is not easy to get the vortex bound states analytically. However, we can understand the problem in another way. The Hamiltonian for a general SC in the basis $\Psi^\dag({\bf k})=(c_{\uparrow}^\dag({\bf k}),c_{\downarrow}^\dag({\bf k}),c_{\downarrow}(-{\bf k}),-c_{\uparrow}(-{\bf k}))$ has the form
\begin{eqnarray}\label{Hsc_general}
H_{sc}&=&\left(\begin{array}{cc}
          H_0(k_\parallel,k_z)-\mu     &     \Delta                        \\
          \Delta^\dagger               &     \mu-H_0(k_\parallel,k_z)      \\
          \end{array}
          \right),
\end{eqnarray}
where $H_0(k_\parallel,k_z)$ is the normal state Hamiltonian and $\mu$ is the chemical potential. Apparently, $k_z$ can be renormalized into the coefficients in the Hamiltonian and $H_{sc}$ has no essential differences between the $k_z=0$ and $k_z\neq 0$ case. Therefore, the Berry phase criterion must also be true when $k_z\neq0$, which means that the DSs must have VPTs at each $k_z$-point in between the two Dirac points while the STIs has no robust VPTs at $k_z\neq 0$. Since the $Z_2^{0D}$ topological invariant is no longer well defined at $k_z\neq 0$, the VPTs must corresponds to a change of the $Z^{0D}$ topological number in the main text.

On the other hand, as the topological property is always protected by a finite energy gap, a small perturbation which does not break the symmetry of the system can not lead to a topological phase transition. As a result, the VPTs at $k_z=0$ and $k_z\neq 0$ must occur in the same subspace of the rotational symmetry. Specifically, the VPT at each $|k_z|<\sqrt{\frac{m}{t_3}}$ for the DS in Eq.\ref{TDSC_4} is attributed to two states with angular momentum $+1$ and $-1$, and this contribute a robust nodal phase. For the STI in Eq.\ref{STIC_4}, since the VPT is contributed by two states with angular momentums $0$, they can hybridize with each other at an arbitrary $k_z$ and this results in a full-gap phase in general.

It can be inferred from the above analysis that, the energy spectrum of the vortex line in doped superconducting STIs are always full-gap, since the VPTs must occur in the subpace with angular momentums $0$. This stems from the fact that, the basis of a STI must contain two Kramers' doublets with the same angular momentum.

Based on the above analysis, we summarize the possible topological VPTs in doped superconducting DSs with $C_n$-rotational symmetry in Table.\ref{VPT_general_DS}. In the calculations, we start from the DSs with continuum rotational symmetry and treat the warping terms which break the continuum rotational symmetry into $C_n$-rotational symmetry as perturbations.

%

\begin{table}[bt]
\caption{\label{VPT_general_DS} In the weak pairing limit, if the warping terms which break the continuum rotational symmetry into $C_n$-rotational symmetry are treated as perturbations, we summarize the subspaces (represented by the eigenvalues) of the $C_n$-rotational symmetry where the VPTs occur for different kinds of DSs in section C.}
\begin{ruledtabular}
\begin{tabular}{c|c|c|c|c|c}
               & $H_{3T}$  & $H_{3N}$  & $H_{4T}$  & $H_{4N}$   & $H_{6T}^{\uppercase\expandafter{\romannumeral1}}$ \\
 \colrule
 subspace    &   $e^{i\frac{2\pi}{3}}\oplus e^{i\frac{4\pi}{3}}$   &   $e^{i\frac{2\pi}{3}}\oplus e^{i\frac{4\pi}{3}}$, $1$
               &   $e^{i\frac{\pi}{2}}\oplus e^{i\frac{3\pi}{2}}$    &   $e^{i\frac{\pi}{2}}\oplus e^{i\frac{3\pi}{2}}$, $1$  &   $e^{i\frac{\pi}{3}}\oplus e^{i\frac{5\pi}{3}}$  \\
  \hline\hline
               & $H_{6N}^{\uppercase\expandafter{\romannumeral1}}$  & $H_{6T}^{\uppercase\expandafter{\romannumeral2}}$  & $H_{6N}^{\uppercase\expandafter{\romannumeral2}}$  & &   \\
 \colrule
 subspace    &    $e^{i\frac{\pi}{3}}\oplus e^{i\frac{5\pi}{3}}$, $1$
               &    $e^{i\frac{2\pi}{3}}\oplus e^{i\frac{4\pi}{3}}$        &    $e^{i\frac{2\pi}{3}}\oplus e^{i\frac{4\pi}{3}}$, $e^{i\frac{\pi}{3}}\oplus e^{i\frac{5\pi}{3}}$
               &        &        \\
\end{tabular}
\end{ruledtabular}
\end{table}

\section{Effects of IS perturbation on the VPTs}

Since such kind of perturbation does not break the rotational symmetry, it can not destroy the nodal phase. On the other hand, for a SC, the PHS, $P$, and the IS, $I$, demands
\begin{eqnarray}\label{symmetry_IS_break}
PH_{sc}({\bf k})P^{-1}=-H_{sc}(-{\bf k}), \quad
IH_{sc}({\bf k})I^{-1}=H_{sc}(-{\bf k}),
\end{eqnarray}
based on which we can know that the degenerate points are no longer necessary to occur at zero energy in the superconducting vortex line.

\end{widetext}

\end{document}